\documentclass[prb,aps,twocolumn,amsmath,amssymb,floatfix,showpacs,
superscriptaddress]{revtex4}

\usepackage[dvips]{graphics}
\usepackage{color}

\begin{document}

\title{Interplay of magnetic field and geometry in magneto-transport of 
mesoscopic loops with Rashba and Dresselhaus spin-orbit interactions}

\author{Shreekantha Sil}

\affiliation{Department of Physics, Visva-Bharati, Santiniketan, West
Bengal-731 235, India}

\author{Santanu K. Maiti}

\email{santanu@post.tau.ac.il}

\affiliation{School of Chemistry, Tel Aviv University, Ramat-Aviv,
Tel Aviv-69978, Israel} 

\author{Arunava Chakrabarti}

\affiliation{Department of Physics, University of Kalyani, Kalyani,
West Bengal-741 235, India}

\begin{abstract}

Electronic transport in closed loop structures is addressed within a 
tight-binding formalism and in the presence of both the Rashba and 
Dresselhaus spin-orbit interactions. It has been shown that any one of 
the spin-orbit (SO) fields can be estimated precisely if the other one is 
known, by observing either the transmission resonance or anti-resonance 
of unpolarized electrons. The result is obtained through an exact analytic 
calculation for a simple square loop, and through a numerically exact 
formulation for a circular ring.  The sensitivity of the transport 
properties on the geometry of the interferometer is discussed in details. 

\end{abstract}


\maketitle

\section{Introduction}

Spintornics is a recent field of utmost research interest that include 
magnetic memory circuits, quantum computers~\cite{zutic,datta,ando,aharony1} 
magnetic nano-structures and quasi one-dimensional semiconductor rings which 
have been acknowledged as ideal candidates for investigating the effects of 
quantum coherence in low-dimensions, and have been examined as the 
prospective quantum devices~\cite{moldo,engels,vlaminck,ando1}. A central 
mechanism governing the physics in the meso- and nano-length scales are 
the Rashba and Dresselhaus spin-orbit interactions which result from a 
structural inversion asymmetry~\cite{rashba}, and  bulk inversion 
asymmetry~\cite{dresselhaus,meier} respectively. The effects are pronounced 
in quantum rings formed at the interface of two narrow gap semiconductors, 
as already discussed in the literature~\cite{koga,premper}.  

Needless to say, an accurate estimation of the spin-orbit interaction (SOI)
strengths is crucial in the field of spintronics. The Rashba spin-orbit
interaction (RSOI) can be controlled by a gate voltage placed in the 
vicinity of the sample~\cite{premper,cmhu,grundler} and hence, can be 
`measured'. Comparatively speaking, reports on the techniques of 
measurement of the Dresselhaus spin-orbit interaction (DSOI) are relatively 
few~\cite{premper,cmhu,grundler}. Very recently we have put forward an idea
of estimating the DSOI strength provided the RSOI is known by measuring
a minimum in the Drude weight~\cite{santanu}. A minimum in the Drude weight
appears only when the strengths of the RSOI and DSOI are exactly identical.

From a closer look at the Rashba and Dresselhaus SO interactions it turns out 
that both the interactions are equivalent to the $SU(2)$ gauge field which 
introduces a phase in the wave function. In this communication we describe
the role of this phase and considering its effect on quantum interference
we develop a simple idea about how one of the two SOI's can be estimated 
while the other is known. This quantum interference effect in presence of
SO interactions has not been described in our previous work~\cite{santanu}
and the present analysis may provide a basic understanding of designing
switching devices for spintronic applications in near future. 
A simple version of a quantum ring, in the form of a loop with a rhombic 
geometry is considered for an analytical attack on the problem, while 
numerically exact results are provided for circular rings with and without 
disorder and with a magnetic flux threading these polygonal structures. We 
adopt a tight-binding formalism in contrast to a recently proposed scheme 
where a continuous version of the model is presented~\cite{ramaglia} to 
consider the combined effect of the RSOI and the DSOI. When strengths of 
\begin{figure}[ht]
{\centering \resizebox*{6.5cm}{3.5cm}{\includegraphics{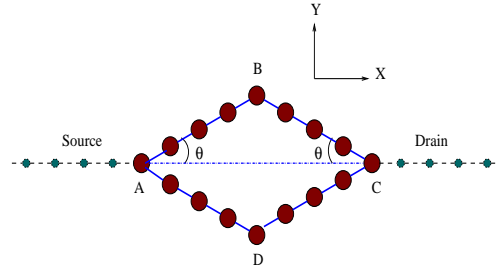}}\par}
\caption{(Color online). Schematic view of a mesoscopic square loop 
subjected to RSOI and DSOI and connected to the leads (source and drain) 
at its two extremities.}
\label{square}
\end{figure}
the RSOI and DSOI are equal, the end to end transmission across the rhombic 
loop is shown to be equal to unity and vanishes when the loop is threaded by 
a magnetic flux $\phi$ equal to the half flux-quantum ($\phi_0/2$). Thus, one 
can estimate the DSOI by observing the peak when $\phi=0$ or dip when 
$\phi=\phi_0/2$ in the transmission (conductance) spectrum when the RSOI 
is known, and vice versa.  

The idea is extended to the case of rings with circular geometry, where we 
have evaluated the transmission coefficient numerically. The essential 
differences in the cases of a rhombic loop and a circular ring are discussed 
to highlight the sensitivity of the results on the loop geometry.

In what follows we describe the procedure and the results. In section II, 
we present the model and the method. In section III we discuss the 
sensitivity of the results on the geometry of the closed loop structures, 
and in section IV we draw our conclusions.

\section{The model and the method}

\noindent
{\bf The Hamiltonian:} Let us consider the rhombic loop depicted in 
Fig.~\ref{square} which is threaded by a magnetic flux $\Phi$. Each side 
of the loop is of length $L$, and the loop contains $N$ number of 
equispaced atomic sites with `lattice constant' $a$. Within a tight-binding 
framework the Hamiltonian for this network in the presence of the RSOI and 
DSOI reads,
\begin{equation}
\mbox{\boldmath $H$} = \mbox{\boldmath $H_0$} -i\alpha \mbox{\boldmath 
$H_R$} + i\beta \mbox{\boldmath $H_D$}
\end{equation}
where,
\begin{equation}
\mbox{\boldmath $H_0$} = - \sum_{i} \left(\mbox{\boldmath $c_{i}^{\dagger} 
t$} \mbox{\boldmath $c_{i+1}$} e^{i\phi} + \mbox{\boldmath $c_{i+1}^{\dagger} 
t$} \mbox{\boldmath $c_{i}$} e^{-i\phi} \right)
\label{ham0}
\end{equation}
The Rashba and Dresselhaus spin-orbit parts of the Hamiltonian, viz, 
\mbox{\boldmath $H_R$} and \mbox{\boldmath $H_D$}, are given by, 
\begin{eqnarray}
\mbox{\boldmath $H_R$}  & = &   
\sum_{i} \left(\sin\theta\mbox{\boldmath $c_{i}^{\dagger}$} 
\mbox{\boldmath$\sigma_x$} \mbox{\boldmath $c_{i+1}$} \right.
- \left. \cos\theta\mbox{\boldmath $c_{i}^{\dagger}$}
\mbox{\boldmath$\sigma_y$}  
\mbox{\boldmath $c_{i+1}$} \right) e^{i\phi} + h.c. \nonumber \\
\mbox{\boldmath $H_D$}  & = &   
\sum_{i} \left(-\cos\theta \mbox{\boldmath $c_{i}^{\dagger}$}
\mbox{\boldmath$\sigma_x$}   
\mbox{\boldmath $c_{i+1}$} \right. + \left. 
\sin\theta \mbox{\boldmath $c_{i}^{\dagger}$}
\mbox{\boldmath$\sigma_y$}  
\mbox{\boldmath $c_{i+1}$}\right) e^{i\phi} + h.c. \nonumber \\
\label{hamiltonian}
\end{eqnarray}
where, $i$ refers to the atomic sites in the arms of the loop.
$\phi = 2 \pi \Phi/N\phi_0$, and $\phi_0 = hc/e$, the fundamental 
flux-quantum. The other operators in Eq.~\ref{hamiltonian} are as follows.\\
\mbox{\boldmath $c_{i}$}=$\left(\begin{array}{c}
c_{i, \uparrow} \\
c_{i, \downarrow}\end{array}\right) \hskip 0.2cm \mbox{and} \hskip 0.2cm
$ \mbox{\boldmath $t$}=$t\left(\begin{array}{cc}
1 & 0 \\
0 & 1 \end{array}\right)$. \\
~\\
\noindent
Here the site energy of an electron at any $i$-th site is assumed to 
be zero throughout the geometry. $t$ is the nearest-neighbor hopping 
integral. $\alpha$ and $\beta$ are the isotropic nearest-neighbor transfer 
integrals which measure the strengths of Rashba and Dresselhaus SOI, 
respectively.  
\mbox{\boldmath $\sigma_x$},
\mbox{\boldmath $\sigma_y$} and \mbox{\boldmath $\sigma_z$} are the Pauli
spin matrices. $c_{i, \sigma}^{\dagger}$ ($c_{i,\sigma}$) is the creation
(annihilation) operator of an electron at the site $i$ with spin 
$\sigma$ ($\uparrow,\downarrow$).

\noindent
{\bf Eigenvalues and eigenfunctions of the Hamiltonian:} These are obtained 
by adopting a $\bf{k}$-space description of the Hamiltonian given in 
Eq.~\ref{hamiltonian}, viz, ${\bf H} = \sum_{k} {\bf c_k^\dag H_k c_k}$. 
Using discrete Fourier transform 
$c_k = \frac{1}{\sqrt{N}} \sum_n c_n \exp(-i {\bf k}.n{\bf a})$, 
the Hamiltonian matrix reads, 
\begin{equation}
\mbox{\boldmath $H_k$}=\left(\begin{array}{cc}
\epsilon_k & \gamma_k + i \delta_k \\
\gamma_k - i \delta_k & \epsilon_k 
\end{array}\right) \\
\label{hmatrix}
\end{equation}
where, 
\begin{eqnarray}
\epsilon_k & = & -2 t \cos(ka + \phi) \nonumber \\
\gamma_k & = & 2\left(\alpha \sin \theta + \beta \cos \theta\right) 
\sin(ka + \phi) \nonumber \\
\delta_k & = & 2\left(\alpha \cos \theta + \beta \sin \theta\right) 
\sin(ka + \phi)\nonumber  
\end{eqnarray}
While writing the above expressions, we have set the lattice constant $a = 1$.
The energy eigenvalues are obtained from Eq.~\ref{hmatrix}, and are given by, 
$E_{k_{\pm}} = \epsilon_k \pm \sqrt{\gamma_k^2 + \delta_k^2}$. 

Let us denote the left vertex of the loop in Fig.~\ref{square} as the 
`origin' $(0,0)$. A simple but lengthy algebra now allows one to write the 
wave function for an energy $E$ at a distance $L$ along any arm of the 
rhombic loop as, 
\begin{equation}
|\Phi_E(L,a) \rangle = \mathcal R_E(L,\alpha,\beta,\theta) |\Phi_E(0)\rangle
\label{wavefunction}
\end{equation}
where, the elements of the matrix $\mathcal R_E(L,\alpha,\beta,\theta)$ are,
\begin{eqnarray}
\mathcal R_E(L,\alpha,\beta,\theta)_{11} &=& \frac{1}{2}\left(e^{ik_+La} + 
e^{ik_{-}La}\right) \nonumber \\
\mathcal R_E(L,\alpha,\beta,\theta)_{12} &=& \frac{1}{2}
\left(e^{i(k_{+}La+\nu_{k+})} - e^{i(k_{-}La+\nu_{k-})}\right) \nonumber \\
\mathcal R_E(L,\alpha,\beta,\theta)_{21} &=& \frac{1}{2}
\left(e^{i(k_{+}La-\nu_{k+})} - e^{i(k_{-}La-\nu_{k-})}\right) \nonumber \\
\mathcal R_E(L,\alpha,\beta,\theta)_{22} &=& \frac{1}{2}\left(e^{ik_+La} + 
e^{ik_{-}La}\right)
\label{Rmatrix}
\end{eqnarray}
Here, $\nu_{k\pm} = \tan^{-1}(\delta_{k\pm}/\gamma_{k\pm})$. $k_{\pm}$ are 
the wave vectors corresponding to the energy values 
$\epsilon_k \pm \sqrt{\gamma_k^2 + \delta_k^2}$, as already mentioned.
\vskip 0.25cm
\noindent
{\bf Transmission of unpolarized electrons:}
Let us now assume that the electrons enter the loop at the point $A$ 
through the source, and are drained out at $B$. In addition, without any 
loss of generality, and to get a relatively simple set of equations, we 
take our loop to be a square one with $\theta = \pi/4$. For an electron 
traveling in the loop in the clockwise sense with a specified energy $E$, 
the wave vector $k_{\pm}$ are the solutions of the equation 
$E = E_{k_{\pm}}$, and can be explicitly written as, 
\begin{eqnarray}
k_{\pm}a & = & \cos^{-1}\xi_{\pm}(E) - \phi 
\end{eqnarray}
where, 
\begin{eqnarray}
\xi_{\pm}(E) & = & \frac{1}{\sqrt{1 + \frac{(\alpha+\beta)^2}{t^2}}} 
\left[- \frac{E}{2t \sqrt{1 + \frac{(\alpha+\beta)^2}{t^2}}} \right.
\nonumber \\ 
 & \pm & \left. \frac{\alpha+\beta}{t} \sqrt{1 -\frac{E^2}{4t^2 
(1 + \frac{(\alpha+\beta)^2}{t^2}}}\right]
\end{eqnarray}
In a similar manner, we need to work out the wave vectors $k'_{\pm}$ for 
the electrons with the same energy $E$, and traveling in the 
counter-clockwise sense. The result is, 
\begin{eqnarray}
k'_{\pm}a & = & \cos^{-1}\xi'_{\pm}(E) - \phi 
\end{eqnarray}
where, 
\begin{eqnarray}
\xi'_{\pm}(E) & = & \frac{1}{\sqrt{1 + \frac{(\alpha-\beta)^2}{t^2}}} 
\left[- \frac{E}{2t \sqrt{1 + \frac{(\alpha-\beta)^2}{t^2}}} \right.
\nonumber \\
 & \pm & \left. \frac{|\alpha-\beta|}{t} \sqrt{1 - \frac{E^2}{4t^2 
(1 + \frac{(\alpha-\beta)^2}{t^2}}}\right]
\label{equ10}
\end{eqnarray}
The probability that the electrons travel in the clockwise or the 
counter-clockwise sense are assumed to be equal. The transmission amplitude 
is given by the matrix, 
\begin{equation}
\tau = \left(\begin{array}{cc}
\tau_{\uparrow\uparrow} & \tau_{\uparrow \downarrow}\\
\tau_{\downarrow \uparrow} & \tau_{\downarrow \downarrow} 
 \end{array}\right) 
\label{trans1}
\end{equation}
which, for $\theta = \pi/4$, simplifies to, 
\begin{eqnarray}
\tau & = & \frac{1}{2} \left[\mathcal R_E(L,\alpha,\beta,-\pi/4) 
\mathcal R_E(L,\alpha,\beta,\pi/4) \right. \nonumber \\
 & & + \left. \mathcal R_E(L,\alpha,\beta,\pi/4) 
\mathcal R_E(L,\alpha,\beta,-\pi/4)\right]
\label{trans2}
\end{eqnarray}
$\mathcal R_E$ matrices can be cast in to the forms, 
\begin{eqnarray}
\mathcal R_E(L,\alpha,\beta,\pi/4) & = & \frac{e^{i\phi L}}{2} 
\left(\mathcal A_1 I + \frac{\mathcal B_1 \sigma_x}{\sqrt{2}} - 
\frac{\mathcal B_1 \sigma_y}{\sqrt{2}}\right) \nonumber \\
\mathcal R_E(L,\alpha,\beta,-\pi/4) & = & \frac{e^{i\phi L}}{2} 
\left(\mathcal A_2 I + \frac{\mathcal B_2 \sigma_x}{\sqrt{2}} + 
\frac{\mathcal B_2 \sigma_y}{\sqrt{2}} \right) \nonumber \\
\label{trans3}
\end{eqnarray}
with, 
\begin{eqnarray}
\mathcal A_1 & = & \exp(ik_{+}La) + \exp(ik_{-}La) \nonumber \\
\mathcal A_2 & = & \exp(ik_{+}'La) + \exp(ik_{-}'La) \nonumber \\ 
\mathcal B_1 & = & \exp(ik_{+}La) - \exp(ik_{-}La) \nonumber \\ 
\mathcal B_2 & = & \exp(ik_{+}'La) - \exp(ik_{-}'La)
\end{eqnarray} 
The transmission amplitude given by Eq.~\ref{trans2} can now
be explicitly written as, 
\begin{eqnarray} 
\tau & = & \frac{1}{8} \left[(2\mathcal A_1 \mathcal A_2 + \sqrt{2} 
(\mathcal A_2 \mathcal B_1 + \mathcal A_1 \mathcal B_2) \sigma_x 
\right. \nonumber \\ 
& + & \left. \sqrt{2} (\mathcal A_1 \mathcal B_2 - \mathcal A_2 
\mathcal B_1) \sigma_y )\right. \cos 2\phi L \nonumber \\ 
 & + & \frac{i}{4} (\sigma_x\sigma_y - \sigma_y\sigma_x) \mathcal B_1 
\mathcal B_2 \sin 2\phi L
\label{trans4}
\end{eqnarray}  
The coefficient of transmission for an incoming {\it up-spin} electron is 
$T_\uparrow = |\tau_{\uparrow \uparrow} + \tau_{\downarrow \uparrow}|^2$, 
and the transmission coefficient for an incoming {\it down-spin} electron is 
$T_\downarrow = |\tau_{\downarrow \downarrow} + \tau_{\downarrow \uparrow}|^2$, 
so that the final transmission coefficient for spin unpolarized electrons 
turns out to be, 
\begin{eqnarray} 
T & = & \frac{1}{2} (T_\uparrow + T_\downarrow) \nonumber \\
& = & \frac{1}{2} \left [\frac{|\mathcal A_1 \mathcal A_2|^2}{8} + 
\frac{|\mathcal A_2\mathcal B_1 + \mathcal A_1 \mathcal B_2|^2}{16} + 
\frac{|\mathcal A_1\mathcal B_2 - \mathcal A_2 \mathcal B_1|^2}{16}\right] 
\nonumber \\
& = & \cos^2\left (\frac{(k'_+ - k'_{-})}{2}La \right ) + 
\sin^2\left (\frac{(k_{+} - k_{-})}{2}La
\right) 
\nonumber \\
& &
\times \cos^2\left (\frac{(k_{+} - k_{-})}{2}La \right), 
\hskip 0.5cm \mbox{for} \hskip 0.15cm \phi=0 \nonumber \\
& =& 4 \sin^2\left (\frac{(k_+ - k_{-})}{2}La \right)
\sin^2\left (\frac{(k'_+ - k'_{-})}{2}La \right), \nonumber \\
& & \mbox{for} \hskip 0.15cm \phi=\pi/4L
\label{trans5}
\end{eqnarray}
When $\alpha=\beta$, from Eq.~\ref{equ10} we observe that 
$\xi'_+ = \xi'_{-}$ or $k'_+ = k'_{-}$ and it gives 
$\mathcal B_2=0$. Therefore, for $\phi=0$, the transmission coefficient $T=1$,
and, $T=0$ for $\phi=\pi/4L$. Thus we get perfect transmission for $\phi=0$
while a vanishing of transmission coefficient for $\phi=\pi/4L$.

\section{Effect of loop geometry}

Electronic transport turns out to be sensitive to the geometry of the 
mesoscopic loop. To this end, we have numerically calculated the 
\begin{figure}[ht]
{\centering \resizebox*{6.5cm}{2.75cm}{\includegraphics{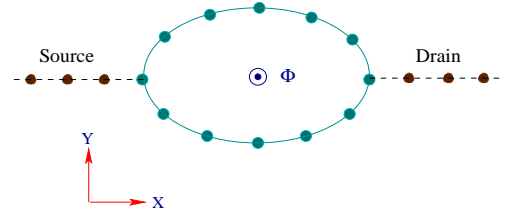}}\par}
\caption{(Color online). Schematic view of a mesoscopic ring subjected 
to RSOI and DSOI and threaded by a magnetic flux $\Phi$. The ring is
symmetrically connected to the leads (source and drain).}
\label{ring}
\end{figure}
two-terminal spin transport in a ring geometry threaded by a magnetic flux 
$\Phi$ (Fig.~\ref{ring}). The role of $\Phi$ will be discussed later. 
Here, for the time being, we ignore the flux. 
\begin{figure}[ht]
{\centering \resizebox*{7.25cm}{3.5cm}{\includegraphics{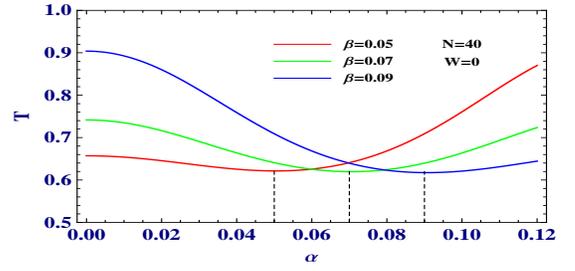}}\par}
\caption{(Color online). Two-terminal transmission coefficient of a 
$40$-site ordered ring for different values of the RSOI ($\alpha$) and 
the DSOI ($\beta$). $\Phi$ is set at $0$.}
\label{order}
\end{figure}
\begin{figure}[ht]
{\centering \resizebox*{7.25cm}{3.5cm}{\includegraphics{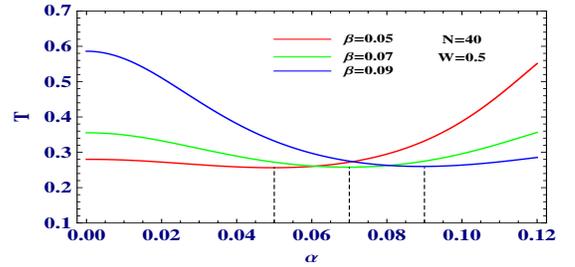}}\par}
\caption{(Color online). Two-terminal transmission coefficient of a 
$40$-site disordered ring for different values of the RSOI ($\alpha$) and 
the DSOI ($\beta$). The results have been averaged over $60$ disorder 
configurations. $\Phi$ is fixed at $0$.}
\label{disorder}
\end{figure}
For a ring like structure, the azimuthal angle keeps on changing as one 
traverses the perimeter of the ring. This generates an effective site 
dependent hopping integral in the Hamiltonian~\cite{santanu}. As a result, 
scattering takes place as the electron travels across the sites on the ring. 
The scattering becomes a maximum when $\alpha = \beta$~\cite{santanu}, and 
naturally, the two-terminal transport is expected to show up a minimum at 
$\alpha = \beta$. We use exact numerical methods. In Fig.~\ref{order} we 
show the variation of the two-terminal transmission 
\begin{figure}[ht]
{\centering \resizebox*{7.25cm}{3.5cm}
{\includegraphics{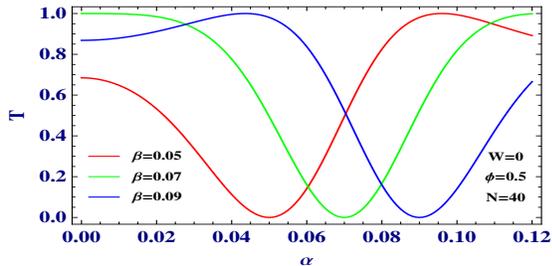}}\par}
\caption{(Color online). Two-terminal transmission coefficient of a $40$-site 
ordered ring for different values of RSOI ($\alpha$) and the DSOI ($\beta$).
$\Phi$ is set equal to $0.5$.}
\label{orderwithflux}
\end{figure}
\begin{figure}[ht]
{\centering \resizebox*{7.25cm}{3.5cm}
{\includegraphics{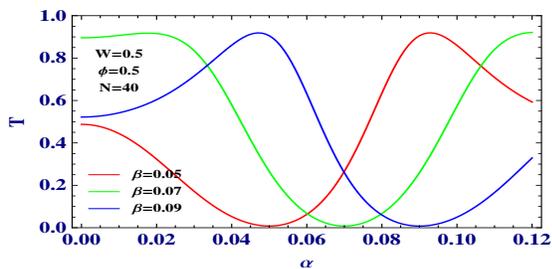}}\par}
\caption{(Color online). Two-terminal transmission coefficient of a 
$40$-site disordered ring for different values of the RSOI ($\alpha$) and 
the DSOI ($\beta$). The results have been averaged over $60$ disorder 
configurations. $\Phi$ is fixed at $0.5$.}
\label{disorderwithflux}
\end{figure}
coefficient for an ordered ($W=0$, $W$ measures the strength of disorder) 
ring of $40$ sites. Similar observations are 
presented in Fig.~\ref{disorder} for a $40$-site ring with random diagonal 
disorder. The results in the latter case have been averaged over $60$ disorder 
configurations. In both the figures the transmission minimum as $\alpha$ 
equals $\beta$ are obvious. It is interesting to note that the random 
disorder does not destroy the minima, which speaks for the robustness of 
the results from the standpoint of experiments.

Before we end this section, it should be appreciated that, in an 
experiment the transmission minimum is not easy to locate, and hence 
an inaccuracy in the value of the SOI might be introduced. This difficulty 
may be circumvented if the transmission minimum becomes precisely equal to 
zero. This is easily achieved if the ring is threaded by a magnetic flux 
which is set equal to half the flux-quantum $\phi_0 = hc/e$. In 
Figs.~\ref{orderwithflux} and \ref{disorderwithflux} the results are 
presented for a $40$-site ordered ring and a randomly disordered ring of 
the same size. The flux threading the rings is set at $\Phi = \phi_0/2$. The 
two-terminal transmission coefficient exhibits clear zeros in both the cases 
as soon as the DSOI equals the strength of the RSOI. Once again, the 
transmission zero in this special situation is independent of the disorder 
configuration.

\section{Conclusion} 

In conclusion, we have presented exact analytical results to show that 
the spin-orbit interactions present in a mesoscopic sample can be 
measured by observing two-terminal transmission resonance in a simple 
square network. The transmission coefficient is shown to be exactly equal 
to unity when the Rashba and the Dresselhaus interactions become equal in 
strength. Thus, knowing the Rashba interaction, for example, a determination 
of the Dresselhaus term is possible. For a multi-site ring, the transmission 
maximum observed for the single square loop gets converted in to transmission 
minimum. This happens again when the Rashba and the Dresselhaus interactions 
are equal. With a magnetic flux equal to half the flux-quantum, the 
transmission minimum becomes an exact transmission zero, and hence 
facilitates a possible experimental measurement.

\end{document}